\documentclass[letter]{aa} % for the letters 
\usepackage{graphicx}
\usepackage{txfonts}
\begin{document}

   \title{Raman-scattered laser guide star photons to monitor the scatter of astronomical telescope mirrors}
   \titlerunning{Raman-scattered laser guide star photons to monitor the scatter of astronomical telescope mirrors}
   \subtitle{}

   \author{Fr\'ed\'eric P.~A. Vogt
          \inst{1}\thanks{ESO Fellow} \thanks{\email{frederic.vogt@alumni.anu.edu.au}}
          \and
          Jos\'e Luis \'Alvarez \inst{1}
          \and
          Domenico Bonaccini Calia \inst{2}
          \and
          Wolfgang Hackenberg \inst{2}
          \and
          Pierre Bourget \inst{1}
          \and
          %Lodovico Coccato \inst{2} 
          %\and?
          Ivan Aranda \inst{1}
          \and
          Callum Bellhouse \inst{3,1}
          \and
          Israel Blanchard \inst{1}
          \and
          Susana Cerda \inst{1}
          \and 
          Claudia Cid \inst{1}
          \and
          Mauro Comin \inst{2}
          \and
          Marcela Espinoza Contreras \inst{1}
          \and
          George Hau \inst{1}
          \and 
          Pascale Hibon \inst{1}
          \and
          Ronald Holzl\"ohner \inst{2}
          \and
          Yara L. Jaff\'e \inst{4}%$^{\star}$
          \and
          Johann Kolb \inst{1}
          \and
          Harald Kuntschner \inst{2}
          \and
          Pierre-Yves Madec \inst{2}
          \and
          Steffen Mieske \inst{1}
          \and
          Julien Milli \inst{1}
          \and
          Cyrielle Opitom \inst{1}$^{\star}$
          \and
          Diego Parraguez \inst{1}
          \and
          %Claudia Reyes Saez \inst{1}
          %\and
          Cristian Romero \inst{1}
          \and
          Fernando Selman \inst{1}
          \and
          Linda Schmidtobreick \inst{1}
          \and
          Jonathan Smoker \inst{1}
          \and
          Sergio Vera Urrutia \inst{1}
          \and
          Gerard Zins \inst{1}
          }

   \institute{European Southern Observatory (ESO), Av. Alonso de C\'ordova 3107, 763 0355 Vitacura, Santiago, Chile.
                  \and
                 European Southern Observatory (ESO), Karl-Schwarzschild-Str. 2, 85748 Garching, Germany.
                 \and
                 University of Birmingham School of Physics and Astronomy, Edgbaston, Birmingham, England.
                 \and
                 Instituto de F\'isica y Astronom\'ia, Universidad de Valpara\'iso, Avda. Gran Breta\~na 1111, Valpara\'iso, Chile.
                 }

   \date{Received 24.08.2018; accepted 28.09.2018}

% \abstract{}{}{}{}{} 
% 5 {} token are mandatory
 
  \abstract
   {The first observations of laser guide star photons Raman-scattered by air molecules above the Very Large Telescope (VLT) were reported in June 2017. The initial detection came from the Multi-Unit Spectroscopic Explorer (MUSE) optical integral field spectrograph, following the installation of the 4 Laser Guide Star Facility (4LGSF) on the Unit Telescope 4 (UT4) of the VLT. In this Letter, we delve further into the symbiotic relationship between the 4LGSF laser guide star system, the UT4 telescope, and MUSE by monitoring the spectral contamination of MUSE observations by Raman photons over a 27\,month period. This dataset reveals that dust particles deposited on the primary and tertiary mirrors of UT4 -- responsible for a reflectivity loss of $\sim$8\% at 6000\,\AA\ -- contribute ($60\pm5$)\% to the laser line fluxes detected by MUSE. The flux of Raman lines, contaminating scientific observations acquired with optical spectrographs, thus provides a new, non-invasive means to monitor the evolving scatter properties of the mirrors of astronomical telescopes equipped with laser guide star systems.
   }


   \maketitle
%
%-------------------------------------------------------------------

\section{Introduction}
Optical photons propagating through the atmosphere may experience a range of scattering and diffusion mechanisms: some, such as Rayleigh and Mie scattering, are elastic processes. Raman scattering, on the other hand, is an inelastic process, through which a photon looses energy by exciting the ro-vibrational modes of molecules \citep{Herzberg1950,Herzberg1945, Telle2007}. The physics of Raman scattering has been used for many years to study the content and structure of the atmosphere \citep{Leonard1967,Cooney1968,Melfi1969,Cooney1970,Melfi1972,Cooney1972,Penney1976,Keckhut1990,Whiteman1992,Heaps1996,Behrendt2002}. In the astronomical context, Raman scattering associated with the use of a laser guide star system was only recently reported, for the first time, with the Multi-Unit Spectroscopic Explorer \citep[MUSE;][]{Bacon2010} integral field spectrograph at the Very Large Telescope (VLT) on Cerro Paranal \citep{Vogt2017b}, in the wake of the installation of the 4 Laser Guide Star Facility \citep[4LGSF;][]{BonacciniCalia2014a}. Since then, Raman scattering of laser guide star photons has also been observed with the Gran Telescopio Canarias at the Observatorio del Roque de los Muchachos in La Palma \citep{Lombardi2017}, with the Gemini South telescope on Cerro Pach\'on in Chile \citep{Marin2018}, and with the Subaru telescope on Mauna Kea in Hawaii \citep{Kawaguchi2018}.

In normal operations, neither the laser spots created in the sodium layer at $\sim$90\,km of altitude, nor the bright uplink laser beams, visible through Rayleigh, Mie and Raman scattering, fall within the MUSE Wide Field Mode (WFM) field-of-view of $1^{\prime}\times1^{\prime}$ \citep[see Fig.~2 in][]{Vogt2017b}. One should also note that in the WFM asterism, the uplink laser beams are \emph{diverging}, and never pass directly above the primary mirror. Although pictures of the 4LGSF in operation tend to visually suggest otherwise, the laser guide stars are located $\sim$62$^{\prime\prime}$($\equiv27$\,m at 90\,km of altitude) away from the center of the MUSE field-of-view, to be compared with the location of the laser launch telescopes, 5.51\,m away from the center of the primary mirror. These facts imply the existence of one (or more) mechanism(s) responsible for bringing the laser photons, scattered while on their way up to the sodium layer, into the scientific field-of-view of MUSE. 

In this Letter, we identify one of these mechanisms -- scattering off dust particles on the primary and tertiary mirrors of UT4 -- and demonstrate that it is responsible for (60$\pm$5)\% of the Raman-scattered photons measured by MUSE in WFM, when the reflectivity of the primary mirror is reduced by $\sim$8\% at 6000\,\AA. We do so via dedicated monitoring observations of the laser lines in MUSE WFM observations, which form the direct continuation of the series of experiments presented by \cite{Vogt2017b}. In the remainder of this Letter, all wavelengths are quoted in air, unless explicitly mentioned otherwise. Whenever we refer to \textit{laser photons}, we mean photons that were originally emitted by the laser guide star system, irrespective of whether they were subsequently Rayleigh-scattered, Mie-scattered, Raman-scattered, or absorbed and re-emitted by a sodium atom.

\section{Observations and data reduction}

The MUSE observations that we are considering here consist of several single 60\,s, 120\,s, 180\,s or 300\,s exposures. During each exposure, the telescope is tracking and guiding on an empty field: ``empty'' in the sense that the field is chosen to contain no entry in the USNO-B1 catalogue \citep[complete down to V=21\,mag;][]{Monet2003}. For each MUSE exposure, the guide star asterism is the nominal WFM-AO one: a square with the laser guide stars located 62$^{\prime\prime}$ away from the field center. These observations were first acquired in the last commissioning run dedicated to the 4LGSF system (stand-alone), and were subsequently performed after each installation of a new component of the Adaptive Optics Facility \citep[AOF;][]{Arsenault2013}, including the Deformable Secondary Mirror \citep[DSM;][]{Arsenault2006a,Briguglio2014} and GALACSI \citep[][]{Stuik2006,LaPenna2016}. Sequences were also acquired after the recoating of the primary and tertiary mirror of UT4 in mid-2017, and in mid-2018 following the installation of NanoBlack\texttrademark\ foil -- initially over a first and subsequently over all the telescope spiders -- to mitigate the so-called \emph{low-wind effect} for MUSE Narrow Field Mode observations \citep[][]{Milli2018}. 

We stress here that the same observation procedure was followed each time. In particular, the 4LGSF system was in all cases operated manually, in a standalone mode which makes MUSE entirely oblivious to it. The DSM was operated in static (no-AO) mode, even after the installation of the GALACSI AO module. The only exceptions are a) one MUSE exposure within the 2017-04-15 sequence acquired with the AO loop closed for comparison purposes (no difference in the flux of the Raman lines was observed), and b) the observations acquired on 2017-08-15. The latter observations were acquired during the afternoon, with the telescope in parked position, and the dome closed. In this specific case, no active optics correction could be applied to the primary mirror, and the exact size (and shape) of the 4LGSF asterism is uncertain. 

All the individual exposures were fetched from the ESO archive, and reduced using the MUSE pipeline 2.2.0 \citep[][]{Weilbacher2015} via \textsc{reflex v2.8.5} \citep{Freudling2013}. We do not perform any telluric subtraction, nor any sky subtraction. All the data was processed using a unique set of master calibrations from late December 2017, which implied the assembly of a custom-build \textsc{Reflex} workflow to do so. The set of master calibrations includes a master bias, a master lamp flat, a master twilight flat, a master wavelength calibration, a master line spread function, a master geometry frame, and a master instrument response. 

The following reasons motivate our decision to employ a unique set of master calibrations (including a master response function) for the data reduction, rather than a mixed set. First, all the MUSE observations presented in this Letter were acquired during commissioning or technical nights, and were not all calibrated with the same level of precision. For example, illumination corrections (useful to account for the temperature fluctuations within the instrument throughout a night) or flux standard stars were/could not be acquired in all cases. Second, the availability of daily calibrations varies from sequence to sequence. Third, the transmission of the telescope improved with the recoating of its primary and tertiary mirrors (``M1'' and ``M3'') in August 2017, but since dust particles on these mirror are responsible for a large part of the contamination of the MUSE exposures by laser photons (see below), the flux of laser lines \textit{decreased} post-recoating. 

All these reasons lead to hard-to-quantify variations in the quality of local calibrations for each observing sequence. Given the overall stability demonstrated by the MUSE instrument over the years, and the fact that all our observing nights were subject to clear or photometric conditions (with the exception of 2018-07-19 and 2018-09-13 which had thin cirrus), we thus opt for a single set of master calibrations. We estimate that this choice can lead to a mismatch in the derived line fluxes at a 5\% level for exposures acquired on a given night, and at an additional 5\% level for exposures acquired on different nights. We note that these error levels are always smaller than the signals we will discuss in the next Section.

\section{Results}

\begin{figure*}[htb!]
\centerline{\includegraphics[scale=0.5]{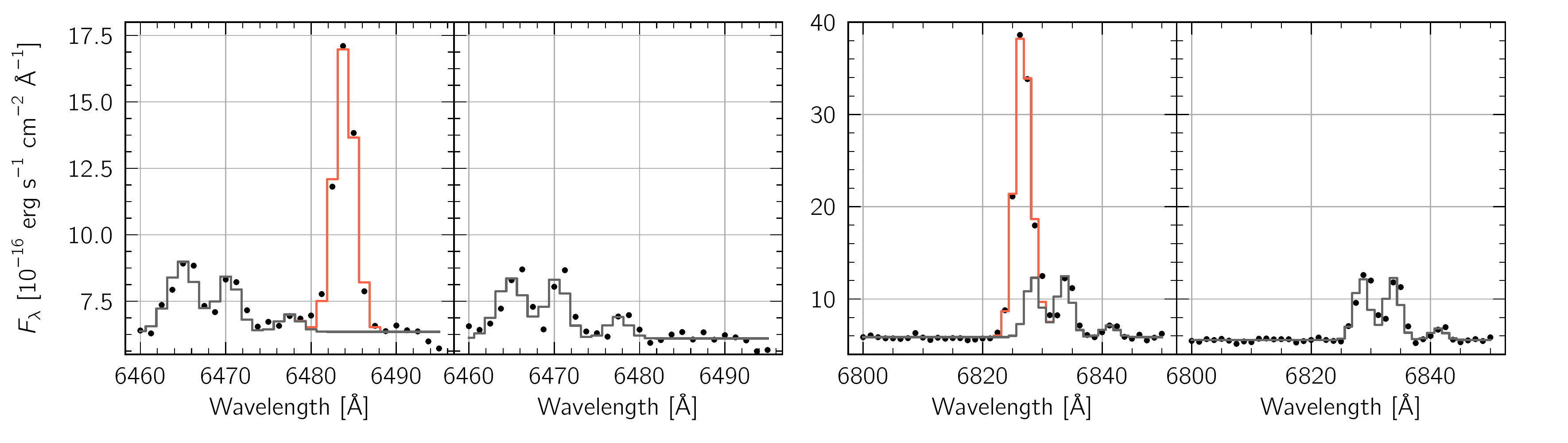}}
\caption{Left: comparison of the spectral fit of the O$_2$ Raman line region from two consecutive MUSE exposures, with and without laser propagation. The black dots trace the spectral flux density $F_\lambda$ integrated within the central 30$^{\prime\prime}$ of the MUSE field-of-view. The dark grey line denotes the sky and background components (only) of the fit. The full fit (that includes the Raman line) is in red. For the observations acquired with no laser propagation, the flux of the Raman lines is fixed at 0. Right: idem, but for the spectral region around the N$_2$ Raman line.}\label{fig:fit}
\end{figure*}

For every MUSE exposure, we extract the integrated spectra from a central circular region of the reduced datacube,  $30^{\prime\prime}$ in diameter. We then measure the associated flux of the main laser line \citep[at 5889.959\,\AA, the residual signal visible through the MUSE notch filter;][]{Vogt2017b}, of the N$_2(\nu_{1\leftarrow0})$ Raman vibrational line at 6827.17\,\AA, and of the O$_2(\nu_{1\leftarrow0})$ Raman vibrational line at 6484.39\,\AA. The latter two lines consist in fact of the entire Q-branch \citep{Herzberg1950}, spectrally unresolved by MUSE. For simplicity, we simply refer to them as the N$_2$ and O$_2$ Raman lines in the remainder of this article. The line fluxes are derived from the least-square fit of a model composed of a single gaussian component (for the laser lines), a constant background level, and single Gaussian components for the near-by sky lines, over spectral windows of 90\,\AA, 50\,\AA, and 35\,{\AA} wide for the 5889.959\,\AA, N$_2$ and O$_2$ lines, respectively. We voluntarily fix the dispersion to 1.2\,{\AA} for all the lines, and the wavelength of the sky lines to 6828.45\,\AA, 6833.65\,\AA, 6841.24\,\AA\ (near the N$_2$ Raman line) and 6484.39\,\AA, 6465.3\,\AA, 6470.6\,\AA, 6477.5\,\AA\ (near the O$_2$ Raman line). This approach ensures robust and consistent results, even for the frames with lower signal-of-noise in the laser lines. The chosen integration aperture of 30$^{\prime\prime}$ in diameter ensures that the N$_2$ and O$_2$ lines are always detected with S/N$\geq$25. Representative examples of the fitting outcome for the O$_2$ and N$_2$ lines are presented in Fig.~\ref{fig:fit}, for data acquired on 2017-04-16.

\begin{figure}[htb!]
\centerline{\includegraphics[scale=0.5]{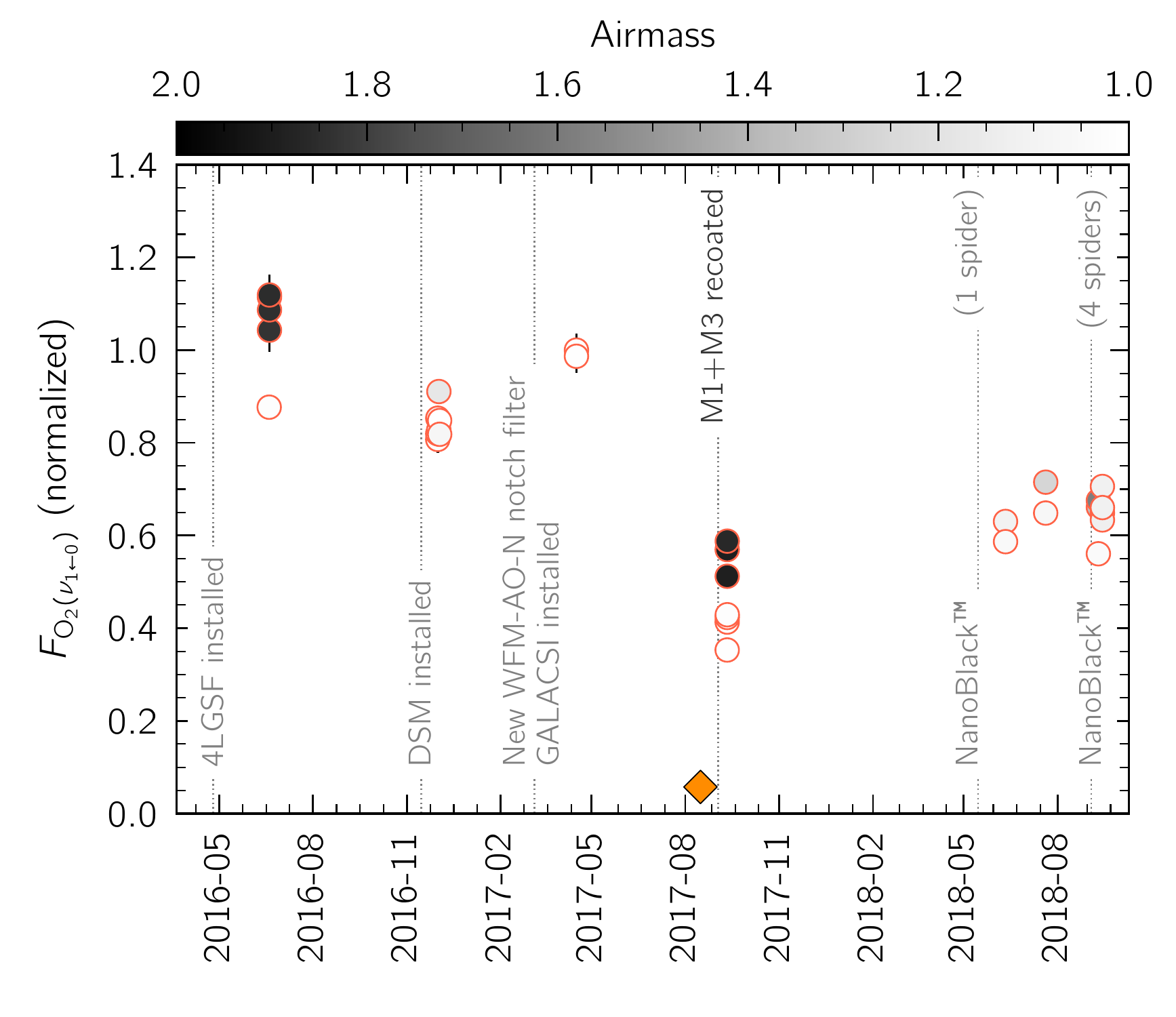}}
\caption{O$_2$ line fluxes $F_{\text{O}_2(\nu_{1\leftarrow 0})}$ extracted from MUSE observations of empty sky fields, acquired with the nominal WFM asterism of laser guide stars (62$^{\prime\prime}$ in radius), normalized to the pre-recoating flux level. Each symbol, color-coded as a function of the airmass of the observation, corresponds to a single exposure. The timeline of specific events affecting UT4 are all indicated. The yellow diamond corresponds to the exposure acquired with the telescope in parked position, and the four 22\,Watt lasers propagating in the closed dome on 2017-08-15 (see Fig.~\ref{fig:in-dome}).}\label{fig:data}
\end{figure}

The derived O$_2$ line fluxes $F_{\text{O}_2(\nu_{1\leftarrow 0})}$ for all observations, normalized to the measurement of (33.7$\pm$0.9)$\times$10$^{-16}$ erg\,s$^{-1}$\,cm$^{-2}$ acquired on 2017-04-16, are presented as a function of time in Fig.~\ref{fig:data}. Each symbol, color-coded as a function of the airmass of the observation, corresponds to a single exposure. The error bars (often smaller than the symbols) indicate the 3-$\sigma$ fitting uncertainty only, and do not include the error associated with using a uniform set of master calibrations. For simplicity, we focus our attention on the O$_2$ Raman line as a proxy for all the laser lines because a) it is detected with a larger S/N than the main laser line at 5889.959\,\AA\ (seen through the notch filter), and b) its fitting is not strongly affected by underlying skylines, unlike the N$_2$ Raman line (see Fig.~\ref{fig:fit}). We also note that the flux of the main laser line at 5889.959\,{\AA} and the N$_2$ Raman line have a behavior qualitatively identical to the O$_2$ Raman line.

\begin{figure}[htb!]
\centerline{\includegraphics[width=\columnwidth]{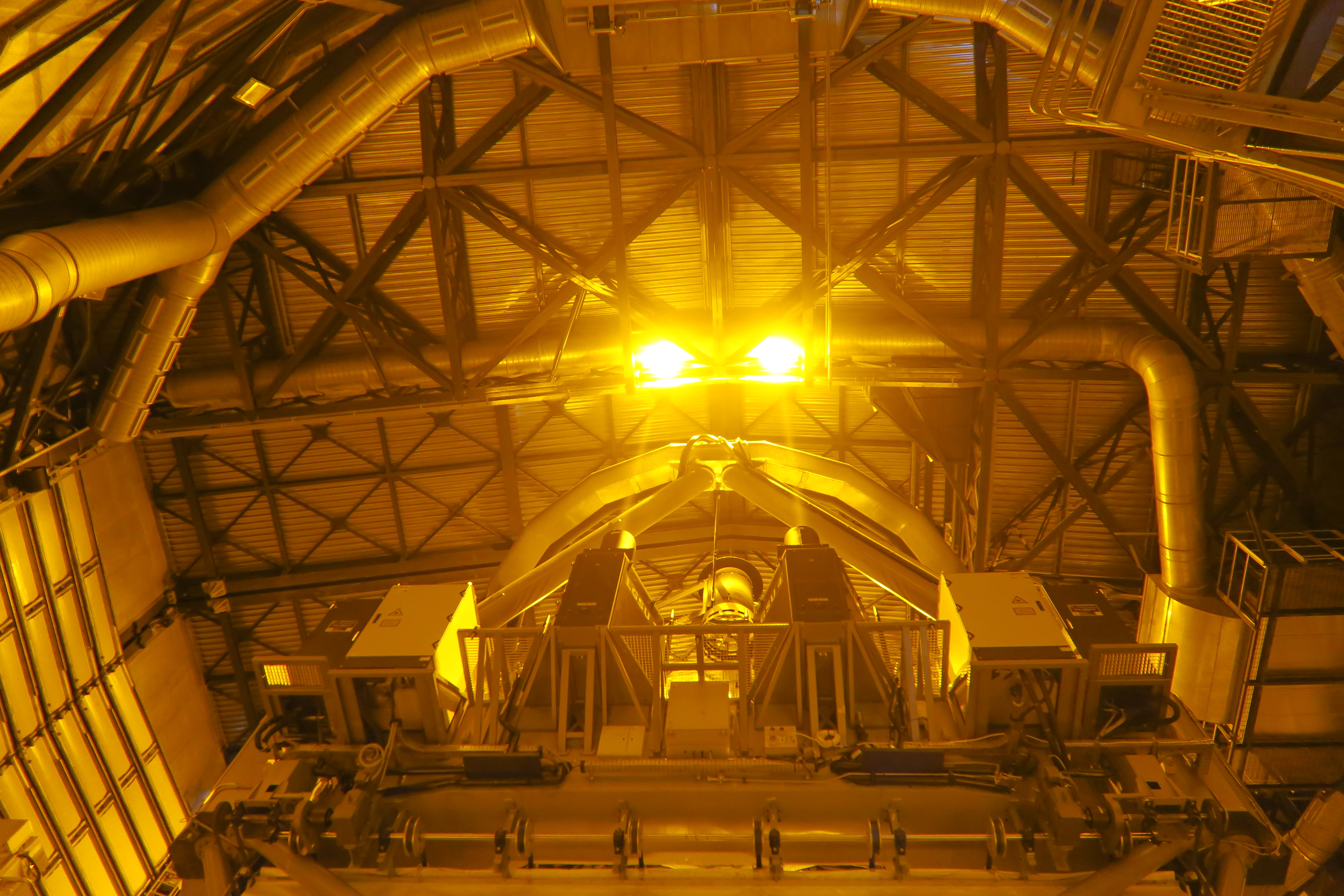}}
\caption{Picture of the interior of UT4 acquired on 2017-08-15, with the dome closed, the telescope parked, and all four lasers from the 4LGSF propagating. LGS1 and LGS4 are visible in the picture, with their beam hitting the cylindrical structure of an air conditioning duct on the dome roof. The overall glow in the picture is evidence for the high amount of scattering of laser photons by the dome surface. All upper and lower lights in the dome were switched off during the test.}\label{fig:in-dome}
\end{figure}

The asterism size associated with the observations performed on 2017-08-15 is uncertain: it most certainly was not a perfect square. The very nature of this observation, performed with the dome closed, prevented us from confirming the asterism shape and size using the Laser Pointing Camera \citep[][]{BonacciniCalia2014,Centrone2016}. We can only be certain that the laser launch telescopes must have been pointing somewhere within their allowed mechanical range of 440\arcsec. This range is equivalent to a maximum positional uncertainty of $\sim$6\,cm at 30\,m of height, implying that within the volume of the UT4 dome, the position of the laser uplink beams during the closed-dome observations is still very consistent with the open-dome cases. 

The close-dome exposure reveals that the intensity of the main 5889.959\,{\AA} laser line seen by MUSE is 1800$\pm$120\,\% stronger than with the dome open. This is not surprising: as illustrated in Fig.~\ref{fig:in-dome}, the laser beams directly hit the reflective metallic surface of the dome structure in this configuration, leading to a very high amount of reflections of laser photons within the dome volume. Despite these numerous reflections on the dome surfaces, the integrated travel path of the laser photons in the air inside the dome is not sufficient to reach the open-dome flux level of the N$_2$ and O$_2$ Raman lines: as illustrated in Fig.~\ref{fig:data}, the intensity of the Raman lines in the close-dome exposure is $6.0\pm0.5$\,\% of the maximum, pre-recoating, open-dome exposure. This close-dome test thus indicates clearly that the majority of the laser photons detected by MUSE in normal operations originate from outside of the dome environment.

Next, we turn our attention to the open-dome observations. We find no correlation between the flux of the laser lines and the air pressure on the ground, the relative humidity at 30\,m above ground, the precipitable water vapour, or the density of 0.5\,$\mu$m and 5\,$\mu$m dust particles at 20\,m above ground at the time of the observations. Exposures acquired pre- and post-DSM installation are largely consistent with one another, indicating that replacing the original secondary mirror of UT4 with the new DSM did not affect the mechanism(s) responsible for bringing laser photons inside of the MUSE WFM field-of-view. 

The same cannot be said for the recoating of M1 and M3 that took place in late August 2017. The recoating process led to a decrease of (60$\pm$5)\% of the intensity of the laser lines seen by MUSE, with respect to pre-recoating levels. 

The primary and tertiary mirrors of UT4 are normally recoated every 18 months \citep{Giordano2004}, but this recoating was the first one since August 2009. This delay was driven by a general maintenance need of the Paranal recoating plant, together with a problem with the grain size of the sputtering target. The plant returned to full operations in mid-2016. Between August 2009 and August 2017, M1 and M3 underwent monthly CO$_2$ cleanings, and one full manual wash in September 2013. Following its 2017 recoating, the reflectivity of the primary mirror at 6240\,\AA\ increased from 81\% to 90\%, whereas its scatter decreased from 15.3\% to 0.3\%, measured using a CT-7 reflectometer-scatterometer in a handful of zones of 6\,mm in diameter.  The positive impact of the recoating is also confirmed by the throughput improvement seen by MUSE measured from the observations of spectrophotometric standard stars, and summarized in Table~\ref{table:through}.

\begin{table}[htb!]
\center
\footnotesize
\caption{Improvement in the total throughput of MUSE (instrument+telescope) following the recoating of the primary and tertiary mirror of UT4 in August 2017.}\label{table:through}
\begin{tabular}{c c c }
\hline
Wavelength & Throughput (pre-) & Throughput (post-) \\
\AA & \% & \%  \\
\hline
\hline
5000 & $22\pm1$ & $28\pm1$  \\
6000 & $28\pm1$ & $35\pm2$ \\
7000 & $33\pm1$ & $40\pm2$  \\
8000 & $25\pm1$ & $30\pm2$  \\
9000 & $15\pm1$ & $18\pm1$  \\
\hline
\end{tabular}
\end{table}

We explain the decrease in the intensity of the laser lines in MUSE exposures post-recoating with the removal of dust particles from the surface of M1 and M3. A fraction of the Rayleigh-, Mie- and Raman- scattered laser photons generated in the laser up-link beams -- primarily in the initial, densest 35\,km of the atmosphere -- naturally fall onto the surface of these mirrors: initially, with an angle such that they do not find themselves heading towards the MUSE WFM field-of-view. Dust particles on the surface of M1 and M3 can however act as a scattering source, imparting an angular kick to the laser photons, redirecting some of them into the scientific field-of-view \citep{Elson1979,Spyak1992,Spyak1992a, Holzlohner2017}. As the scattering occurs on the surface of mirrors in the optical path, no amount of baffling downstream from M3 can stop the scattered laser photons from reaching the MUSE detectors: a fact consistent with the lack of difference in the Raman line fluxes pre- and post-GALACSI installation. 

The fact that the scattering properties of the primary mirror of UT4 decreased to $<1\%$ post-recoating indicates that the laser photons still detected by MUSE immediately after the recoating must enter its field-of-view through a distinct set of mechanisms, that remain formally unidentified at the time of publication of this Letter. The microroughness of the mirror surface, which ought to dominate the clean-mirror scatter at optical wavelengths \citep{Spyak1992a}, is among the plausible culprits. For any given state of the primary mirror, we note that the intensity of the laser lines detected by MUSE is also airmass dependent: their flux increases by (30$\pm$5)\% at an airmass of 2, when compared to an airmass of 1 (see Fig.~\ref{fig:data}). 

In the 13 months following the recoating, the intensity of the laser lines detected by MUSE has increased by $\sim2\pm1$ \% per month (with respect to the maximum line flux measured prior to the recoating), in spite of the monthly CO2 cleanings. This is not surprising, since CO2 cleaning \citep{Zito1990} is known a) to have difficulties removing 50-100 $\mu$m particles \citep{Kimura1995}, and b) to be somewhat less efficient when performed with a baseline longer than 1-2 weeks \citep{Toomey1994}. Our measurements imply that one would reach a pre-recoating Raman contamination level of MUSE exposures in a total of 30$\pm$7 months: a duration somewhat shorter than the 4 years separating the 2017 recoating from the preceding full manual wash of 2013. The atmospheric aerosol content on Cerro Paranal is however known to be subject to strong seasonal variations \citep{Giordano1994}, and the increase of the mirror scatter is likely not a linear process. Measurements of the intensity of the Raman lines in MUSE WFM-AO observations with a higher temporal frequency than presented in this Letter is thus essential to enable more accurate long-term predictions of the M1 and M3 scatter.

Altogether, our dedicated set of MUSE observations acquired alongside the 4LGSF system over a 27 month period reveal the clear impact of dust on the primary and tertiary mirrors of UT4. The Raman lines in MUSE WFM-AO observations, unaffected by the notch filter, appear as an excellent non-invasive means to characterize, in a qualitative sense, the evolution of the scattering properties of M1 and M3 over time. These lines are now being fitted within the updated MUSE data reduction pipeline for AO observations, initially as part of the correction of the sky background. As such, they can be monitored on any night when MUSE WFM-AO observations are acquired. In practice, this corresponds to a monitoring frequency of a sub-hourly cadence on a quasi-nightly basis, given the robustness of the entire AOF installation in terms of operations. This  frequency could allow, for example, to assess the efficiency of the regular CO2 cleanings, and could ease the implementation of a condition-based mirror washing/recoating. This ``cost free'' (time-wise) and non-invasive approach to monitor the scatter of telescope mirrors may also prove to be of use for other facilities using a laser guide star system feeding an optical spectrograph: for example, the Extremely Large Telescope, to be equipped with up to eight 22\,Watt Na lasers to feed corrected wavefronts to the HARMONI optical-infrared integral field spectrograph \citep[][]{Thatte2010,Thatte2014}. 

\begin{acknowledgements}
We are grateful to Angela Cortes, Fran\c{c}ois Rigaut and C\'eline d'Orgeville for enlightening discussions, Claudia Reyes Saez for her support operating the 4LGSF in standalone mode, Philippe Duhoux and Stefan Stroebele for their support during the observations acquired in September 2017, and Lodovico Coccato for his help setting up a custom \textsc{Reflex} workflow to reduce MUSE data using master calibration files. We thank the anonymous referee for his/her constructive feedback.

This research has made use of \textsc{fcmaker} \citep{Vogt2018b,Vogt2018a}, a \textsc{python} module to create ESO-compliant finding charts for OBs on \textit{p2}. \textsc{fcmaker} relies on \textsc{matplotlib} \citep{Hunter2007}, \textsc{astropy}, a community-developed core \textsc{python} package for Astronomy \citep{AstropyCollaboration2013}, \textsc{astroquery}, a package hosted at \url{https://astroquery.readthedocs.io} which provides a set of tools for querying astronomical web forms and databases \citep{Ginsburg2017}, \textsc{astroplan} \citep{Morris2018}, \textsc{aplpy}, an open-source plotting package for \textsc{python} \citep{Robitaille2012}, and \textsc{montage}, funded by the National Science Foundation under Grant Number ACI-1440620 and previously funded
by the National Aeronautics and Space Administration's Earth Science Technology Office, Computation Technologies Project, under Cooperative Agreement Number NCC5-626 between NASA and the California Institute of Technology. \textsc{fcmaker} uses the VizieR catalogue access tool, CDS, Strasbourg, France. The original description of the VizieR service was published in \cite{Ochsenbein2000}. \textsc{fcmaker} makes use of data from the European Space Agency (ESA) mission Gaia (\url{https://www.cosmos.esa.int/gaia}), processed by the Gaia Data Processing and Analysis Consortium (DPAC, \url{https://www.cosmos.esa.int/web/gaia/dpac/consortium}). Funding for the DPAC has been provided by national institutions, in particular the institutions participating in the Gaia Multilateral Agreement. In particular, \textsc{fcmaker} uses data from the Gaia \citep{GaiaCollaboration2016a} Data Release 2 \citep{GaiaCollaboration2018}. \textsc{fcmaker} also uses data from the Second Digitized Sky Survey (DSS 2). The ``Second Epoch Survey'' of the southern sky was produced by the Anglo-Australian Observatory (AAO) using the UK Schmidt Telescope. Plates from this survey have been digitized and compressed by the STScI. The digitized images are copyright (c) 1993-1995 by the Anglo-Australian Telescope Board. The ``Equatorial Red Atlas'' of the southern sky was produced using the UK Schmidt Telescope. Plates from this survey have been digitized and compressed by the STScI. The digitized images are copyright (c) 1992-1995, jointly by the UK SERC/PPARC (Particle Physics and Astronomy Research Council, formerly Science and Engineering Research Council) and the Anglo-Australian Telescope Board. The compressed files of the ``Palomar Observatory - Space Telescope Science Institute Digital Sky Survey'' of the northern sky, based on scans of the Second Palomar Sky Survey, are copyright (c) 1993-1995 by the California Institute of Technology. All DSS2 material not subject to one of the above copyright provisions is copyright (c) 1995 by the Association of Universities for Research in Astronomy, Inc., produced under Contract No. NAS 5-26555 with the National Aeronautics and Space Administration. This research has also made use of \textsc{mpfit}, a Python script that uses the Levenberg-Marquardt technique \citep{More1978} to solve least-squares problems, based on an original \textsc{fortran} code that is part of the \textsc{minpack}-1 package, of the \textsc{aladin} interactive sky atlas \citep{Bonnarel2000}, of \textsc{saoimage ds9} \citep{Joye2003} developed by Smithsonian Astrophysical Observatory, of NASA's Astrophysics Data System, and of the services of the ESO Science Archive Facility. Based on observations made with ESO Telescopes at the La Silla Paranal Observatory. All the observations described in this article are freely available online from the ESO Data Archive.

Y.~L.~J. acknowledges support from CONICYT PAI (Concurso Nacional de Inserci\'on en la Academia 2017) No. 79170132.
\end{acknowledgements}

\bibliographystyle{aa.bst} % style aa.bst
\bibliography{bibliography_fixed} % your references Yourfile.bib

\end{document}